# One-Dimensional Hole Gas in Germanium/Silicon Nanowire Heterostructures

Wei Lu[†§], Jie Xiang[†§], Brian P. Timko[†], Yue Wu[†] & Charles M. Lieber*[†‡]

[†]Department of Chemistry and Chemical Biology, Harvard University, Cambridge, MA, 02138, USA. [‡]Division of Engineering and Applied Sciences, Harvard University, Cambridge, MA 02138, USA.

[§]These authors contributed equally to this work.

*To whom correspondence should be addressed. E-mail: cml@cmliris.harvard.edu.

___________________________________________________

**Corresponding author:** Charles M. Lieber, Department of Chemistry and Chemical Biology, Harvard University, Cambridge, MA 02138; Tel: 617-496-3169; Fax: 617-496-5442; Email: cml@cmliris.harvard.edu.

# Abstract

Two-dimensional electron and hole gas systems, enabled through band structure design and epitaxial growth on planar substrates, have served as key platforms for fundamental condensed matter research and high performance devices. The analogous development of one-dimensional (1D) electron or hole gas systems through controlled growth on 1D nanostructure substrates, which could open up opportunities beyond existing carbon nanotube and nanowire systems, has not been realized. Here we report the synthesis and transport studies of a 1D hole gas system based on a free-standing germanium/silicon (Ge/Si) core/shell nanowire heterostructure. Room temperature electrical transport measurements show clearly hole accumulation in undoped Ge/Si nanowire heterostructures, in contrast to control experiments on single component nanowires. Low-temperature studies show well controlled Coulomb blockade oscillations when the Si shell serves as a tunnel barrier to the hole gas in the Ge channel. Transparent contacts to the hole gas also have been reproducibly achieved by thermal annealing. In such devices, we observe conductance quantization at low temperatures, corresponding to ballistic transport through 1D subbands, where the measured subband energy spacings agree with calculations for a cylindrical confinement potential. In addition, we observe a "0.7 structure", which has been attributed to spontaneous spin polarization, suggesting the universality of this phenomenon in interacting 1D systems. Lastly, the conductance exhibits little temperature dependence, consistent with our calculation of reduced backscattering in this 1D system, and suggests that transport is ballistic even at room temperature.

Carbon nanotubes and semiconductor nanowires have attracted considerable attention as 1D structures for fundamental studies and also as potential building blocks for nanodevices (1-3). Current synthetic methods can reproducibly yield nanotubes and nanowires with diameters of a few nanometers (4,5), comparable to the de Broglie wavelength of the carriers. In this regime, quantum confinement may affect significantly transport through these materials, thus making them model platforms to study and utilize potentially unique properties of 1D systems (6,7). Indeed, ballistic transport and conductance quantization have been observed in metallic (8) and semiconducting carbon nanotubes (9,10). Similar effects have not been observed in semiconductor nanowires, although the ability to vary size, material composition and electronic properties of semiconductor nanowires in a controlled manner (2,3) offers substantial potential for creating designed 1D systems.

Band structure engineering has been widely used in the past to yield interesting planar systems, including two-dimensional electron gases in GaAs/AlGaAs heterostructures (11) and two-dimensional hole gases in Ge/SiGe heterostructures (12). 2D electron and 2D hole gases have been central to mesoscopic physics (13) and have lead, for example, to the discovery of conductance quantization in quantum point contacts (14) and realization of artificial atoms in quantum dots (15). We use this underlying concept from 2D to design a 1D hole gas system based on a Ge/Si core/shell nanowire heterostructure. Because there is ~500 meV valence band offset (12,16) between the Ge core and Si shell in this heterostructure, free holes will accumulate in the Ge channel when the Fermi level lies below the valance band edge of the Ge core. We have demonstrated that transparent contacts to the hole gas can be achieved as a result of the band line-up, and obtained long carrier mean free path by eliminating scattering from dopants. With these improvements, we have observed ballistic transport through devices up to a few hundred nanometers in length both at low temperature and room temperature.

**Materials and Methods**

**Ge/Si nanowire growth.** Gold nanoclusters of 5 nm, 10 nm and 15 nm diameters (Ted Pella, Inc., Redding, CA) were first deposited on oxidized silicon wafers and placed in a quartz tube furnace. Ge nanowire core growth was initiated by nucleation at 315 °C for 1 minute using 10% $GeH_4$ in $H_2$ (30 sccm) and $H_2$ (200 sccm) at 300 torr, followed by axial elongation at 280 °C for 15 minutes and 100 torr. The i-Si shell was deposited within the same reactor immediately following Ge core growth at 450 °C for 5~10 minutes using $SiH_4$ (5 sccm) at 5 torr. The growth rates for the i-Ge core and i-Si shell were ca. 1 μm/min and 0.5 nm/min, respectively. For the control experiments, i-Ge nanowires were grown using the same procedure as the Ge core, and i-Si nanowires were grown at 435 °C for 20 min using $SiH_4/H_2$ (6/60 sccm) at 60 torr.

**Back-gated devices.** Nanowires were dispersed in ethanol from growth substrates by sonication, and then deposited on degenerately doped Si substrates with 50 nm thermal oxide layer (n-type, resistivity <0.005 Ω-cm, Nova Electronics Materials, Carrollton, TX). Electron beam lithography and metal deposition were used to define Ni source/drain electrodes (50 nm thick). To make contacts to the Ge channel, the devices were annealed at 300 °C for 15 seconds in forming gas ($N_2/H_2$, 90/10%) (Heatpulse 610, Metron Technology, San Jose, CA).

**Top-gated devices.** Atomic layer deposition was used to deposit $Al_2O_3$ dielectric conformally on Ge/Si nanowire devices prepared in the same way as the back-gated devices. Deposition was carried out at 200 °C using a cycle consisting of 1 s water vapor pulse, 2 s $N_2$ purge, 1 s trimethyl aluminum, and 2 s $N_2$ purge; 50 cycles were used to yield a thickness of 6 nm. The top gate was then defined by electron beam lithography, followed by Cr/Au (5/50 nm) deposition.

**Electrical transport measurements.** Room temperature measurements were performed in vacuum at a pressure below $1\times10^{-4}$ torr using a cryogenic probe station (TTP-4, Desert

Cryogenics, Tucson, AZ). Low temperature measurements were carried out using both the probe station and a He-4 cryostat (Oxford Instruments, Abingdon, U.K.). The differential conductance *G* was measured with a lock-in amplifier (SR 830, Stanford Research Systems, Sunnyvale, CA) using an 11 Hz ac excitation of 200 µV superimposed on a dc bias voltage.

**Results and Discussion**

The Ge/Si core/shell (Fig. 1A) nanowires were grown using a reported approach (17), except that both the Ge core and Si shell were not doped. This difference is critical for our studies since it eliminates scattering from ionized dopants in the 2-5 nm thick Si shell adjacent to the Ge channel. A thin Si shell was used in our studies to facilitate electrical contact to the Ge channel, and to reduce the likelihood of dislocations in the shell. The valence band offset of ca. 500 meV between Ge and Si at the heterostructure interface (12,16) serves as a confinement potential for the quantum well, and free holes will accumulate in the Ge channel when the Fermi level lies below the valance band edge of the Ge core (Fig. 1B). High-resolution transmission electron microscopy studies of the Ge/Si nanowires (Fig. 1C) show clearly the core (dark)/shell (light) structure. The lattice fringes and sharp interface between Ge and Si show that the core/shell structure is epitaxial, and is consistent with cross-sectional elemental mapping results. Lower resolution images also indicate that dislocations are not present in these structures. The clean, epitaxial interface in these nanowire heterostructures should produce a smooth confinement potential along the channel. The pseudomorphic strain in the epitaxial core and shell materials is relaxed along the radial direction (17), and this will yield a type II staggered band alignment (12). The light hole and heavy hole bands are expected to split due to the effects of strain and confinement (12).

Electrical transport measurements were made on Ge/Si nanowire devices with lithographically-patterned nickel source/drain electrodes and capacitively coupled back-gate electrodes. A brief annealing process was performed after the source/drain fabrication to facilitate contact to the inner Ge channel. Room-temperature current versus source-drain voltage ($I$-$V_{SD}$) data recorded on a heterostructure with a 10 nm Ge core diameter (Fig. 2A) exhibit a substantial current at zero gate voltage ($V_g = 0$), and a decrease in current as $V_g$ is increased from -10 to 10 V. These results show that the device behaves as a p-type depletion mode field-effect transistor (p-FET), and thus confirm the accumulation of hole charge carriers. Because both the Ge core and the Si shell are un-doped, the existence of a hole gas is a result of the band line-up as illustrated in Fig. 1B, where the Fermi level is pinned below the Ge valance band edge, due to the combined effect of work function, band offset and interface states (18). To further probe this phenomenon, we performed control experiments on intrinsic Si (i-Si) and intrinsic Ge (i-Ge) nanowires. Transport measurements (Fig. 2B) show that both the i-Si and i-Ge nanowires are enhancement mode p-FETs with no carriers at $V_g$=0. The i-Ge nanowire data thus contrast that obtained for the Ge/Si core/shell structure even though the i-Ge nanowires were grown under identical conditions to the Ge core in the heterostructure.

For the Ge/Si core/shell nanowire devices fabricated without the annealing process, carriers need to tunnel through the non-conducting Si layer, which appear as barriers in transport data recorded at low temperatures (inset, Fig. 3A). The Ge/Si devices prepared with tunnel barriers at the contacts show periodic oscillations in $I$ as a function of $V_g$ (Fig. 3A); these are Coulomb blockade oscillations and the device behaves as a single-electron transistor (SET) (19). From the period of the current oscillations in Fig. 3A, we estimate the gate capacitance, $C_g$, to be 3.2 aF for this 112 nm long device. In Fig. 3B, we plot the differential conductance $G = dI/dV_{SD}$ vs. $V_{SD}$ and $V_g$ for the same device. These data exhibit well-defined Coulomb diamonds as expected for transport through a single SET (19). In Fig. 3C we show $G$-$V_{SD}$-$V_g$ for another 385 nm long device with $C_g$ = 15.8 aF, which also shows well-defined Coulomb diamonds characteristics of transport through a single SET. To verify that the tunnel barriers defining the SET are not caused by defects inside the nanowire, which break the nanowire into multiple islands (20), and are often observed in lithographically defined wires (21), we plot (Fig. 3D) the measured gate capacitance $C_g$ as a function of the source-drain separation $L$ (measured from SEM images). Notably, the measured $C_g$ agrees well for $L > 100$ nm with the capacitance calculated using a cylinder-on-plane model $C_g = \dfrac{2\pi\varepsilon\varepsilon_0 L}{\cosh^{-1}(h/r)}$, where $h$ = 50 nm is the oxide thickness, $r$ is the radius of the Ge core, and $\varepsilon$ is the dielectric constant. For $L < 100$ nm, the measured $C_g$ is generally smaller than that predicted by the simple model, and reflects screening by the metal leads when $L$ is comparable to $h$. The scaling of $C_g$ with $L$ is a clear demonstration that the tunnel barriers are formed at the contacts, and importantly, that no significant scattering centers exist inside the channel up to a channel length of at least 500 nm.

In single component semiconductor nanowire devices, a Schottky barrier always forms at the contact since the Fermi level ($E_F$) lies inside the semiconductor band gap (22). In contrast, barriers to the hole gas in Ge/Si nanowires are not intrinsic and can be eliminated, since $E_F$ lies outside the Ge bandgap (Fig. 1B). Indeed, annealing the Ge/Si nanowire devices produces reproducible transparent contacts to the hole gas even at low temperatures. $I$-$V_{SD}$ data obtained at 4.7 K on an annealed device with a 10 nm core (Fig. 4A, inset) close to depletion ($V_g$ = 10 V) are linear around $V_{SD} = 0$, and thus show that the contacts are transparent at low temperatures. We emphasize that the depletion mode p-FET behavior with transparent contacts at both room and low temperatures is observed for essentially all of the Ge/Si nanowire heterostructure devices prepared in this way. The reproducibility of this unique contact behavior demonstrates clearly the impact possible through band structure control.

At small bias the $I$-$V_{SD}$ curves collapse for $V_g$ < 7 V (right inset, Fig. 4A). This behavior is highlighted by plotting $G$ vs. $V_g$ (Fig. 4A), where $G$ first rises sharply and then plateaus for $V_g$ < 7 V. The plateau conductance, ~50 μS, is 0.65 of $2e^2/h$, the value expected for a spin-degenerate single-mode ballistic conductor (13). Variations in the plateau conductance are suggestive of Fabry-Perot interferences (8), but are not quantified here due to their small amplitude. Studies of additional devices show very similar results and highlight the reproducible transport properties of the Ge/Si nanowire system. For example, Fig. 4B shows $G$ vs. $V_g$ recorded at different temperatures for another 10 nm core diameter device. At 4.7 K, the device shows a conductance plateau close to $2e^2/h$, which is consistent with data in Fig. 4A and the value for a single-mode ballistic conductor. Notably, increasing temperature up to 300 K yields little change in the value of the conductance plateau, although the slope becomes somewhat smeared. This fact implies that even at room temperature only a single 1D subband participates in transport and that the mean free path exceeds the channel length of 170 nm; that is, transport through the Ge/Si nanowire remains ballistic up to at least this length scale.

These results contrast data from planar 2DHG devices, where the mobility decreases dramatically with increasing temperature due to scattering with acoustic phonons (23). In a 1D system, acoustic phonon scattering should be suppressed due to the reduced phase space for back-scattering (24). In the Ge/Si nanowires, we use Fermi's golden rule to estimate the acoustic phonon scattering rate as

$$\frac{1}{\tau_{ap}} = \frac{\pi k_B T \Xi^2}{\hbar \rho v_s^2} D(E_F), \tag{1}$$

where $\Xi$ is the deformation potential, $\rho$ is the mass density, and $v_s$ is the velocity of sound. We calculate the density of states $D(E) = \frac{\sqrt{2m^*}}{\pi \hbar \sqrt{E}} \frac{1}{\pi r^2}$ for the first subband using the effective mass for heavy holes, $m^* = 0.28 m_e$ (12,18), where $m_e$ is the free electron mass. We obtain $\tau_{ap} \sim 4.9 \times 10^{-12}$ s at room temperature for a 10 nm core diameter nanowire at $E_F = 10$ meV using an average sound velocity of 5400 m/s and deformation potential of 3.81 eV for Ge (12,25). The mean free path calculated using this value of $\tau_{ap}$ and the Fermi velocity, $v_F = \sqrt{2 E_F / m^*} \sim 1.1 \times 10^5$ m/s, is 540 nm. This estimate is consistent with our experimental results, and moreover, suggests that room-temperature ballistic transport in Ge/Si nanowires is possible on a 0.5 μm scale assuming other scattering processes are suppressed. Room temperature ballistic transport has been reported previously in metallic and semiconducting carbon nanotubes (9,26), and was attributed to the topological singularity at $k = 0$ due to its unique band structure in metallic nanotubes (27,28). Reduced acoustic phonon scattering in 1D explains room temperature ballistic transport in Ge/Si nanowires, although a more detailed theoretical analysis, including the confinement effects of phonon modes, will be needed to quantify the limits of this interesting behavior.

In addition, we have studied devices with a top-gate structure, which increases the gate coupling, in order to probe transport through more than one subband. $G$-$V_g$ data recorded on a 400 nm long device (Fig. 5A) shows four distinct conductance plateaus at 5 K. We attribute these plateaus to transport through the first four subbands in the Ge/Si nanowire, and confirm this assignment by plotting $G$–$V_{SD}$ for different values of $V_g$ (Fig. 5B). In this plot, the conductance plateaus appear as dark regions, labeled as *a-d*, where several $G$-$V_{SD}$ curves at different $V_g$ overlap, since $V_g$ does not affect $G$ in the plateau regions. At large $V_{SD}$ these integer plateaus evolve into "half" plateaus (*f-g*) when the source and drain chemical potentials cross different subbands (29,30). For example, the 0.5 plateau, appearing as the dark region labeled *f*, corresponds to the case where the source potential drops below the first subband bottom while the drain potential still lies above it (29). Similarly, feature *g* corresponds to the 1.5 plateau, which evolves from the second (*b*) and first (*a*) subbands. The small cusp feature near zero-bias in the $G$-$V_{SD}$ data is due to shallow potential barriers (31) with heights of a few meV. The potential barriers are likely caused by non-optimal fabrication process in these top-gated devices, since they are absent in the bottom gated devices studied earlier. The existence of shallow barriers also explains the thermal activation behavior observed in $G$ measured at zero-bias (but not in $G$ measured outside the cusp at 8 mV), and values of the conductance on the plateaus less than multiples of $2e^2/h$.

The assignment of these features to 1D subbands in the Ge/Si nanowires was further analyzed by quantifying the level spacings. Such features appear more pronounced after numerically differentiate $G$ against $V_g$ (29,30). A plot of the transconductance, $dG/dV_g$, as a function of $V_{SD}$ and $V_g$ (Fig. 5C) shows zero or low values at conductance plateaus and high values in the transition regions between plateaus, which are highlighted by dashed lines in Fig. 5C. The subband spacings are obtained directly from these data as the $V_{SD}$ values at the apexes of the full plateaus (i.e., the extrapolated intersections of the dashed lines), which yield $\Delta E_{1,2} = 25$ mV and $\Delta E_{2,3} = 30$ mV. For comparison, we have calculated the subband spacings using an effective mass model with a cylindrical confinement potential with radius *r* to approximate the Ge/Si nanowire structure. The energy levels of the 1D modes due to radial confinement are

$$E = \frac{\hbar^2 u_{ni}^2}{2m^* r^2}, \qquad (2)$$

where $u_{ni}$ is Bessel function $J_n(x)$'s $i^{th}$ zero point, and $m^*$ is the hole effective mass as discussed above. For a nanowire with 14 nm Ge core diameter, we obtain $\Delta E_{1,2} = 25$ mV and $\Delta E_{2,3} = 32$ mV. These values are in good agreement with our experimental data (Fig. 5C), and thus provide strong support for our assignment of discrete 1D subbands in the Ge/Si nanowire heterostructures.

Lastly, a reproducible feature is observed with a conductance value ca. 0.7 times the first plateau in the bottom-gated (Fig. 4, vertical arrows) and top-gated (Fig. 5A, inset; Fig. 5B, label-*e*) devices. Similar features, termed "0.7 structure", have been observed previously on quantum point contacts and quantum wires formed in clean 2DEG samples (29,32,33). This feature is generally believed to be caused by spontaneous spin polarization in low-dimensional interacting electron gas systems (32) due to the formation of a spin gap (33) or a localized spin (29). Temperature dependent data recorded on a back-gated Ge/Si device (Fig. 4B) show that the 0.7 feature initially increases in magnitude and then broadens as temperature is increased to 50 K, consistent with the spin-gap hypothesis (33). These results suggest that the "0.7 structure" is not restricted to electron gas samples, but a universal phenomenon in 1D systems. In this regard, the heavier effective mass of holes in the Ge/Si nanowires compared to electrons will yield a larger interaction parameter (6,34), and should make Ge/Si nanowires interesting to study in greater detail in the future.

In conclusion, we have used band-structure design and controlled epitaxial growth to create a 1D hole gas system in Ge/Si core/shell nanowire heterostructures. We have observed ballistic transport through individual 1D subbands due to confinement of carriers in the radial direction. Significantly, the conductance shows little temperature dependence, suggesting a room temperature carrier mean free path on the order of several hundred nanometers. We believe that the long mean free path, transparent contacts, and high device yield of the Ge/Si system make it attractive for both fundamental studies of strongly interacting low dimensional systems and applications as very high-performance nanoelectronic and quantum devices, and more generally, that band-structure design and controlled epitaxial growth will open many opportunities for fundamental studies in 1D nanowires in the future.

We thank C. Marcus, H. Park and D. Reilly for helpful discussions. C.M.L. acknowledges support of this work by Intel, Defense Advanced Research Projects Agency, and the Army Research Office.

**Figure Captions**

**Figure 1.** Ge/Si core/shell nanowires. Schematics of (A) a cross section through the Ge/Si core/shell structure, and (B) the band diagram for a Si/Ge/Si heterostructure. The dashed line indicates the position of the Fermi level, $E_F$, which lies inside the Si band gap and below the Ge valance band edge. (C) High resolution TEM image of a Ge/Si core/shell nanowire with 15 nm Ge (dark gray) core diameter and 5 nm Si (light gray) shell thickness. The contrast between core and shell is due to difference in atomic weights of Ge and Si. Scale bar is 5 nm.

**Figure 2.** Room temperature electrical transport in Ge/Si nanowires. (A) $I$-$V_{SD}$ characteristics recorded on a 10 nm core diameter Ge/Si nanowire device with source drain separation $L = 1$ µm. The different curves correspond to the back gate voltage $V_g$ values of +10 V (dashed line), 0 (solid line) and -10 V (dotted line). (Lower inset) Schematic of a Ge/Si core/shell nanowire. (Upper inset) $I$-$V_g$ for the same device at $V_{SD} = -1$ V. (B) $I$-$V_{SD}$ measurements on i-Si (blue, 20 nm diameter, 1 µm channel length) and i-Ge (red, 20 nm diameter, 1 µm channel length) nanowires; the data were recorded for $V_g = 0$ and -10 V, corresponding to off and on states, respectively. (Inset) Schematic of the i-Si (blue) and i-Ge (red) nanowires.

**Figure 3.** Coulomb blockade (CB) in unannealed Ge/Si devices. (A) $I$-$V_g$ for a 10 nm core diameter Ge/Si nanowire ($T = 1.5$ K, $V_{SD} = 0.5$ mV, $L = 112$ nm). The gate capacitance is $C_g = e/\Delta V_{CB} = 3.2$ aF, where $\Delta V_{CB}$ is the period of the CB oscillation. Inset, $I$-$V_{SD}$ data taken at $V_g = -9.38$ V showing the CB gap. (B) Coulomb diamonds in $G$-$V_{SD}$-$V_g$ plot for the device in A. (C) $G$-$V_{SD}$-$V_g$ plot for another 10 nm core diameter nanowire with $L = 385$ nm, $C_g = 15.8$ aF for this device. (D) Scaling of $C_g$ with channel length $L$. The red line is the theoretical prediction based on the cylinder-on-plane model discussed in the text. Black diamonds are experimental data from CB measurements.

**Figure 4.** Ballistic transport in Ge/Si 1D hole gas devices. (A) $G$-$V_g$ recorded at 4.7 K for a 10 nm core Ge/Si nanowire with $L$ = 350 nm. (Left inset) $I$-$V_{SD}$ curve recorded at $V_g$ = 10 V. (Right inset) $I$-$V_{SD}$ data recorded at $V_g$ from 10 to -10 V in 1 V steps. (B) $G$-$V_g$ data recorded for another Ge/Si nanowire device ($L$ = 170 nm) measured at different temperatures: black, green, blue, and red correspond to 300, 50, 10, and 4.7 K, respectively. The conductance value on the plateau decreases slightly with decreasing temperature. This reduction is likely due to small potential fluctuations inside the channel, although other factors such as changes in the contact resistance or many-body effects as discussed in Ref. 6 cannot be ruled out. The vertical arrows in (A) and (B) highlight the 0.7 structure.

**Figure 5.** Transport through multiple subbands in a Ge/Si nanowire device. (A) $G$ – $V_g$ recorded at different temperatures for a top-gated device ($V_{SD}$ = 8 mV). The red, blue, green, and black curves correspond to 5, 10, 50, and 100 K, respectively. Data was offset horizontally for clarity. (Upper inset) SEM image of the sample showing the source (S) drain (D) and the top gate (G) electrodes; scale bar is 500 nm. (Lower inset) Zero-bias $G$-$V_g$ recorded at 5 (red), 10 (blue) and 20 (green) K; axes have the same units as the main figure. The vertical arrow highlights the "0.7 structure". (B) $G$-$V_{SD}$ plots recorded at $V_g$ = 0.8 to -3.5V in 50 mV steps with no offset applied ($T$ = 5 K). Data taken in a different thermal cycle from A. Plateaus, labeled *a-g*, which appear dark due to accumulation of $G$ – $V_{SD}$ traces, are discussed in the text. Feature *h* evolves from the first plateau *a* and the 0.7 structure *e*. (C) Transconductance $dG/dV_g$ as a function of $V_{SD}$ and $V_g$ for data in (B). Dashed lines are guides indicating the evolution of conductance modes with $V_{SD}$ and $V_g$. The vertical arrows highlight values of $\Delta E_{1,2}$ and $\Delta E_{2,3}$.

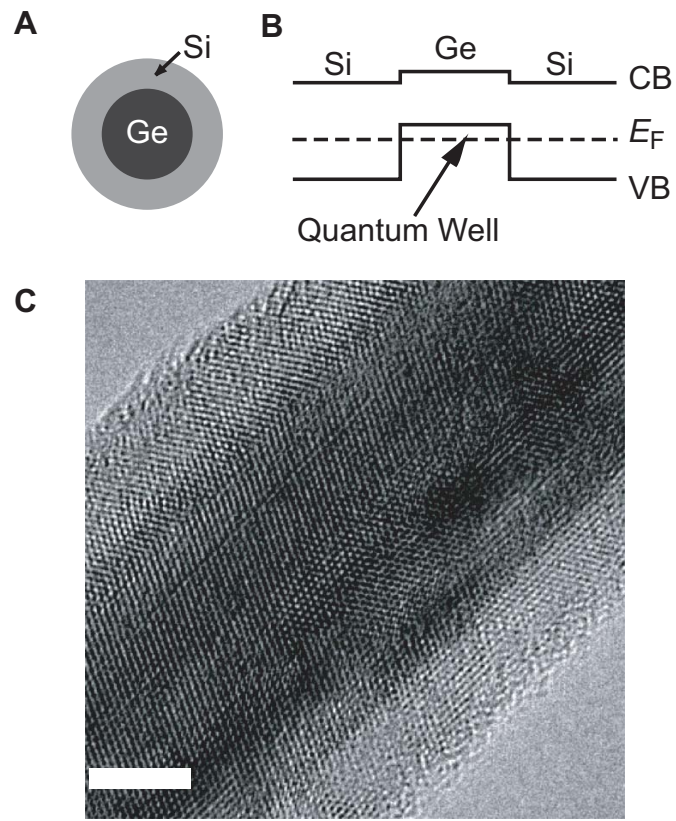

Figure 1

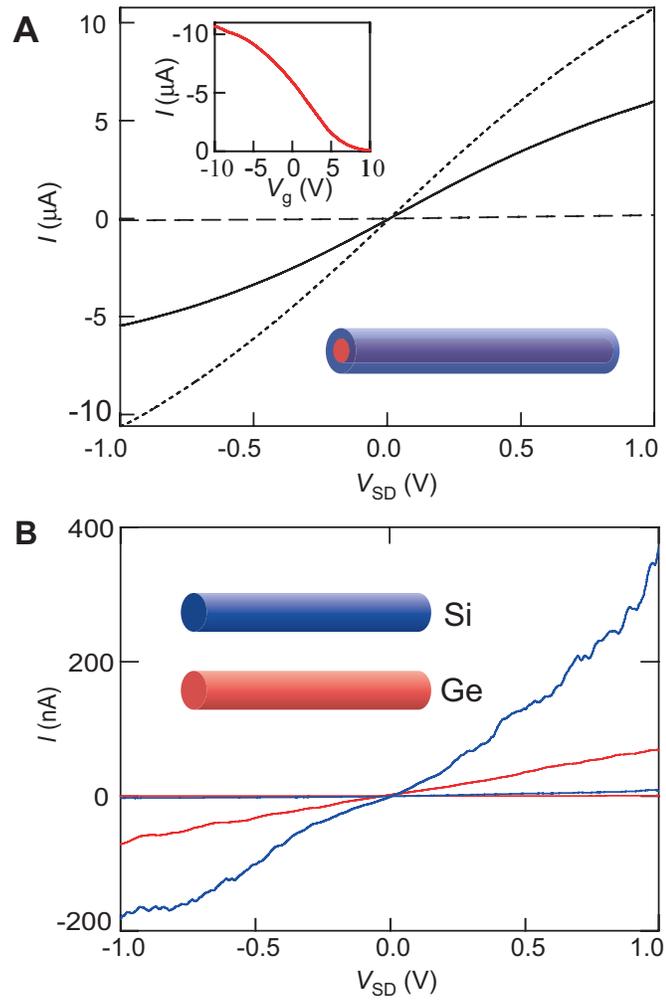

Figure 2

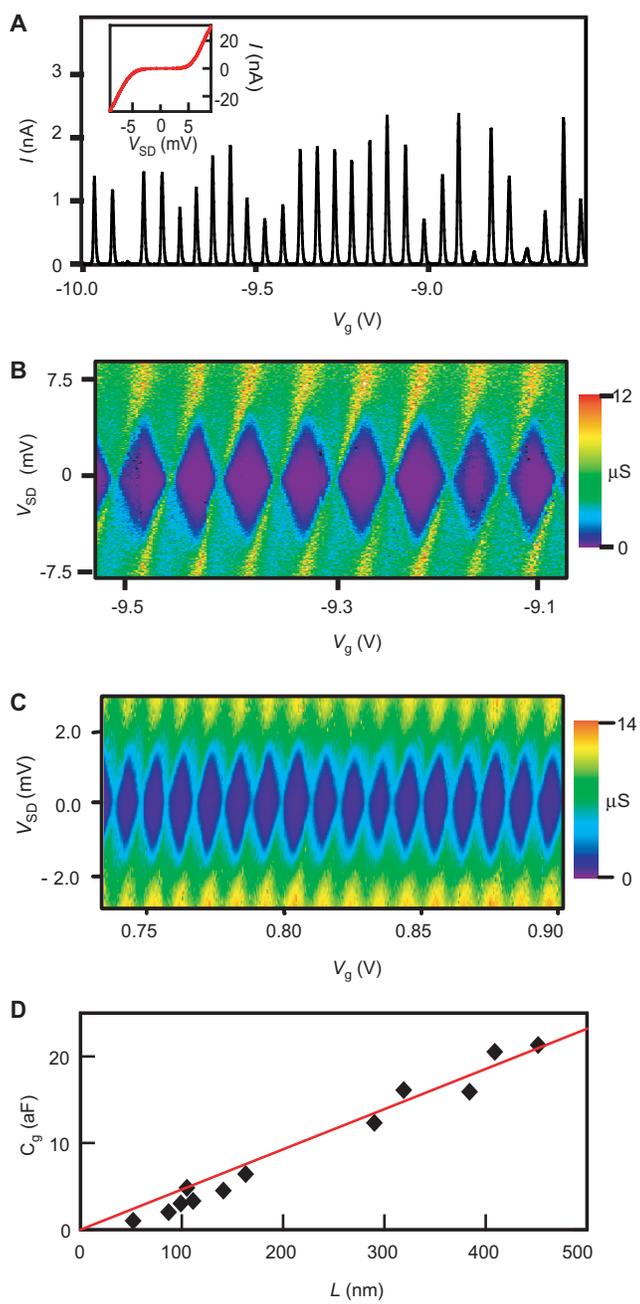

Figure 3

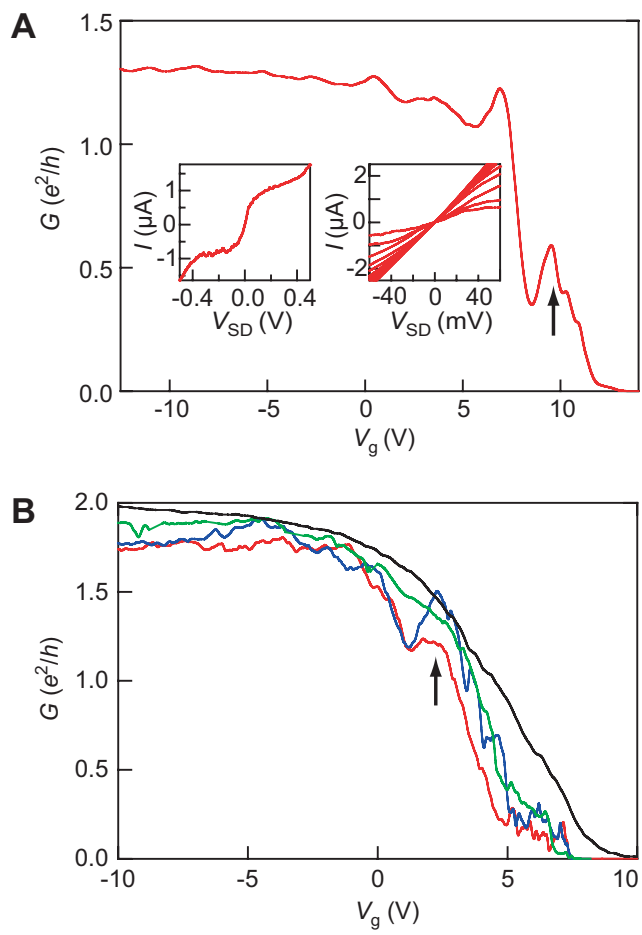

Figure 4

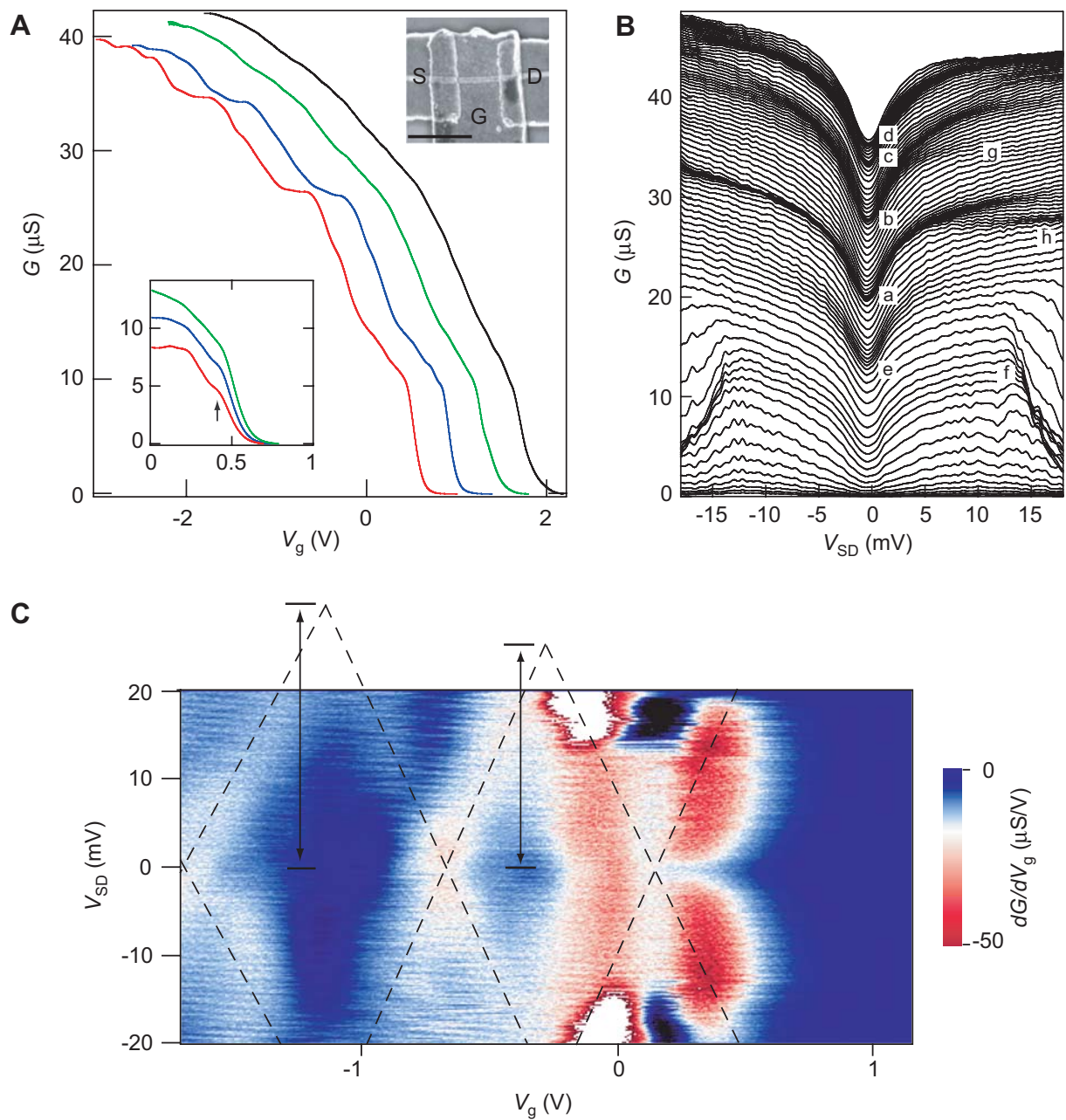

Figure 5